\documentclass[12pt,aps,epsfig,a4paper]{article}
\usepackage[dvips]{epsfig}
\usepackage{amsfonts}
\usepackage{amssymb}
\usepackage{latexsym}
\usepackage{amsmath}
\usepackage{makeidx}

\newcommand{\newc}{\newcommand}

\newc{\ra}{\rightarrow}

\newc{\lra}{\leftrightarrow}
\newcommand{\dis}{\displaystyle}

\newc{\beq}{\begin{equation}}
\newc{\eeq}{\end{equation}}
\newc{\barr}{\begin{eqnarray}}
\newc{\earr}{\end{eqnarray}}

\newcommand{\bvec}[1]{\mbox{\boldmath$#1$}}

\begin{document}

\begin{titlepage}

%%%%%%%%%%%%%%%%%%%%%%%%%%%%%%%%%%%%%%%%%%%%%%%%%%%%%%%%%%%%%%%%%%%%

\title{\large \bf BEATS OF THE MAGNETOCAPACITANCE OSCILLATIONS IN
LATERAL SEMICONDUCTOR SUPERLATTICES}

\author{G. S. Kliros$^{1}$  and P. C. Divari$^{2}$ \\
{\small\it $^{1}$Department of Electronics and Communications
Engineering, }\\ \small\it  Hellenic
Air-Force Academy, Dekeleia Air-Force Base GR-1010, Greece\\
\small\it $^{2}$Department of Physics, University of Ioannina,
Ioannina GR-45110, Greece }

\maketitle

\begin{abstract}
We present calculations on the magnetocapacitance of the
two-dimensional electron gas in a lateral semiconductor
superlattice under two-dimensional weak periodic potential
modulation in the presence of a perpendicular magnetic field.
Adopting a Gaussian broadening of magnetic-field-dependent width
in the density of states, we present explicit and simple
expressions for the magnetocapacitance, valid for the relevant
weak magnetic fields and modulation strengths. As the modulation
strength in both directions increase, beats of the
magnetocapacitance oscillations are observed, in the low magnetic
field range (Weiss-oscillations regime), which are absent in the
one-dimensional weak modulation case.
\end{abstract}

\bigskip

\noindent {\small\it KEYWORDS}: lateral surface superlattices;
density of states; magnetocapacitance.

\bigskip

\noindent

\parskip=0mm

\parindent=1.5em

\end{titlepage}

\section{Introduction}

A lateral surface superlattice (LSSL) combines a system of
nano-fabricated gates with a MODFET, where a periodic pattern is
imposed onto 2DEG at the semiconductor heterostructure\cite{Ross}.
This pattern can be a one- or two-dimensional periodic potential
modulation of different strength allowing for a variety of
artificial periodic structures. The periodic potential is applied
through an array of metal gates whose bias can be
varied\cite{Peet}. The modulation which is purely induced by the
modulated gate potential, leads predominantly to a sinusoidal
potential shape and its strength can be tuned by a nano-patterned
top gate as well as by a back gate electrode\cite{Linde}.
Magnetotransport measurements in LSSL devices have revealed novel
oscillations in the conductivity, the Weiss
oscillations\cite{Weiss} which are superimposed on top of the
well-known Shubnikov-de Haas oscillations. Weiss oscillations were
first discovered in LSSL devices with holographically defined
one-dimensional (1D) superlattices\cite{Winkler}, and were
explained\cite{Been,Gerha} in terms of a commensurability
criterion between the cyclotron orbit diameter at the Fermi level
$2R_c=2\sqrt{2\pi n_e}\hspace{2pt}l^2$ (where $n_e$ is the
electron density and $l=\sqrt{\hbar/eB}$ the magnetic length) and
the superlattice period $\alpha$:
\begin{equation}\label{weiss}
2R_c=(\lambda+{\phi})\alpha \hspace{12pt} \lambda=0,1,2,...
\end{equation}
where $\phi$ is a phase factor.

Besides Weiss oscillations, novel magneto-resistance oscillations
with $1/B$ periodicity have been observed in short-period
superlattices when one flux quantum $h/e$ passes through an
integral number of lattice unit cells\cite{Chow,Wang}. A large
fraction of these studies was performed on square\cite{Schlo} or
rectangular lattices\cite{Albr,Schu}, while only a few experiments
on hexagonal lattices have been published\cite{Nihey,Iye}. For the
majority of the effects, the lattice type is, in principle,
irrelevant. Nevertheless, Altshuler-Aronov-Spivak oscillations
have been observed around zero magnetic field in hexagonal
lattices, while Aronov-Bohm oscillations can be detected in larger
magnetic fields\cite{Iye}.

The capacitance of a mesoscopic structure is of electrochemical
nature. It depends in an explicit manner via the thermodynamical
density of states on the electronic properties of the structure.
The investigation of the magnetic-field dependence of the
electrochemical capacitance plays a significant role in several
experiments, especially those using capacitance spectroscopy.
Capacitive studies of the density of states of a two-dimensional
electron gas are possible because the thermodynamical density of
states appears as a series capacitance to the geometrical
capacitance\cite{Smith}. The behavior of the electrochemical
capacitance can be also be examined from a dynamical point of
view\cite{Buttiker}. At low frequencies, the frequency dependent
capacitance can be expanded in frequency and at the linear order
it is a series combination of the static capacitance and a charge
relaxation resistance. This approach has been performed to the
analysis of mesoscopic capacitors in Refs\cite{Wei,Pomor}.

The oscillatory characteristics caused by commensurability between
the cyclotron diameter $2R_c$ and the modulation period $\alpha$
in a LSSL, reveal themselves in the electrochemical
magnetocapacitance measurements. A pronounced modulation of both
the minima and maxima of the capacitance oscillations has been
observed in LSSLs with 1D weak periodic potential and explained as
a consequence of the oscillatory bandwidth of the
modulation-broadened Landau levels\cite{Weiss}. Although in recent
years special attention has been paid to investigate the magnetic
miniband structure as well as the magneto-transport properties in
a LSSL under two-dimensional weak modulation\cite{Chow,Wang},
magnetocapacitance calculations pertinent to this case are rather
limited\cite{Ismail}.

In this paper we present calculations on the density of states and
the magnetocapacitance of the 2DEG in a LSSL under two-dimensional
weak periodic potential  modulation in the presence of a
perpendicular magnetic field. Adopting a Gaussian broadening of
magnetic-field-dependent width, we present explicit and simple
expressions for the DOS, valid for the relevant weak magnetic
fields and modulation strengths. We investigate the
magnetocapacitance, as a function of the magnetic field, for
various strengths of the periodic potential modulation. In the
low-field range (Weiss-oscillations regime), a beating structure
in the oscillating magnetocapacitance is observed.

The paper is organized as follows: In Section 2 we briefly present
the magnetic spectrum of the 2DEG in a LSSL under two-dimensional
weak periodic potential modulation. In Section 3, the DOS is
calculated analytically and numerically as a function of energy
and magnetic field. In section 4 we present numerical results
concerning the magnetocapacitance oscillations. We conclude with a
summarizing Section 5.

\section{Energy spectrum under weak 2D-periodic modulation potential}

To describe the electrons in the conduction band of a lateral
superlattice at the AlGaAs-GaAs interface in a constant
perpendicular magnetic field $\bvec{B}=B\hat{z}$, we employ a
model of strictly 2DEG with a 2D periodic potential modulation
$V(\bvec{r})$, $\hspace{3pt}\bvec{r}\equiv(x,y)$. Following the
common practice, we adopt an effective mass approximation to take
into account the effect of the crystalline atomic structure over
the charge carriers. The system is described by the well-known
single-particle Hamiltonian  of Bloch electrons in a homogeneous
magnetic field:
\begin{equation}\label{Ham}\mathcal{H}=\frac{1}{2m^\star} \big
(\bvec{p}+e\bvec{A} \big)^2 +V(\bvec{r}) \end{equation} where we
neglect the Zeeman splitting and spin-orbit interactions which are
very small in GaAs systems. The effective mass $m^{\star}$ for
electrons in AlGaAs-GaAs is taken $m^{\star}=0.067~m_e$, where
$m_e$ is the free-electron mass.

We consider a two-dimensional periodic modulation with rectangular
symmetry, i.e:\begin{equation}
\label{poten}V(x,y)=V_x\cos{(\frac{2\pi}{a_x}x)}
+V_y\cos{(\frac{2\pi}{a_y}y)}
\end{equation}
where $V_x$,~$V_y$ are the modulation strengths and $a_x$,~$a_y$
the periods in the corresponding directions. We assume that the
modulation is weak compared to the cyclotron energy $\hbar
\omega_c$ so that mixing of different Landau levels can be
neglected. The homogeneous magnetic field is represented by the
vector potential $\bvec{A}=(-By/2,-Bx/2,0)$.

Applying a direct perturbation theory\cite{Wang} or a projection
operator approach\cite{Klir} to unmodulated system, the energy
spectrum is obtained as a function of the crystal momentum
$(k_x,k_y)$ and the number of flux quanta through a lattice-unit
cell $\Phi/\Phi_0 $, where $\Phi_0 = h/e$ is the magnetic flux
quantum.
\begin{equation}
\label{specrum}
 E(k_x,k_y)= (n+\frac{1}{2}\hspace{3pt})
\hbar\hspace{1pt}\omega_c\hspace{2pt}+V_xF_n(u_x)\cos{(\frac{2a_x}{\pi}u_xk_y)}
 +V_yF_n(u_y)\cos{(\frac{2a_y}{\pi}u_yk_x)}
\end{equation}
where $F_n(u)=\exp{(-u/2)}L_n(u),\hspace{3 pt}
u_x=(\pi\Phi_0/\Phi){a_y}/{a_x},\hspace{3 pt}
u_y=(\pi\Phi_0/\Phi){a_x}/{a_y}$ and $L_n(u)$ are the Laguerre
polynomials. The perturbation approach has been shown to be an
extremely good approximation to the direct diagonalization of the
Hamiltonian, for the parameter values of interest ($V_x, V_y <<
E_F$, where $E_F$ is the Fermi energy) and for magnetic fields
$0.1~T<B<1~T$. Thus, the unperturbed Landau levels broaden into
bands, with width $W_{n}(B)=2(V_x|F_n(u_x)|+V_y|F_n(u_y)|)$, that
oscillates with magnetic field $B$.

In the low-magnetic-field range it is a good approximation to take
the large n-limit of the Laguerre polynomials\cite{Abra} that
appear in $W_{n}(B)$. For a square lattice with $a_x=a_y=a$, the
width of the bandwidth is then obtained as
\begin{equation}
\label{BW}
W_n(B)=\frac{2(V_x+V_y)}{\pi}\sqrt{\frac{a}{R_c}}\hspace{5pt}
\bigg|\cos\bigg(\frac{2{\pi}R_c}{a}-\frac{\pi}{4}\bigg)\bigg|
\end{equation}
where $R_c=l\sqrt{2n_{F}+1}$ is the cyclotron radius with
$n_{F}=E_{F}/\hbar \omega_c$ the Landau index at the Fermi level.
From equation (\ref{BW}) we obtain that the bandwidth reaches zero
(flat band) when $2R_c/a=(\lambda+1/4)$ and maximum when
$2R_c/a=(\lambda+3/4)$ with $\lambda=0, 1, 2, ...$ .

\section{Density of States}

The DOS per unit area of the 2DEG, in the absence of disorder, can
be expressed as a series of $\delta$-functions:
\begin{equation} \label{dosdelta}
D(E)=\frac{g_s}{2\pi l^2} \sum_{nk_xk_y}\delta(E-E_{nk_xk_y})\;
\end{equation}
where $g_s=2$ is the electron-spin degeneracy.

In the case of 1D electrostatic modulation, the $\delta$-functions
in Eq. (\ref{dosdelta}) are broadened into bands with Van Hove
singularities at the edges of each Landau band. On the other hand,
if 2D modulation is imposed, the DOS does not exhibit Van Hove
singularities reflecting the 2D nature of the electron motion in
the Landau bands with finite mean velocities\cite{Klir}.

In practical two-dimensional electron systems, there is always
some broadening present due to scattering centers. We assume a
Gaussian broadening of zero shift and a width $\Gamma$ which,
according to short-range scattering theory of Aoki and
Ando\cite{Aoki}, is proportional to $\sqrt{B}$. In this case Eq.
(\ref{dosdelta}) takes the form
\begin{eqnarray} \label{dosgaus}
\begin{array}{llc}
D(E)&=\dis { \frac{g_s}{2\pi l^2} }
\sum_{nk_xk_y}\frac{1}{\sqrt{2\pi}\Gamma}
\exp{\bigg(-(E-E_{nk_xk_y})^2/2\Gamma^2\bigg)} \\%\nonumber \\
&=\dis {  \frac{g_s}{2\pi^3 l^2} \sum_{n}\int_0^{\pi} } dt_x
\int_0^{\pi} dt_y\frac{1}{\sqrt{2\pi}\Gamma} \exp{\bigg(-(E-E_{n
t_xt_y})^2/2\Gamma^2\bigg)}\;
\end{array}
\end{eqnarray}

We expand the integrand of Eq. (\ref{dosgaus}) in powers of
$\epsilon_{n,t_i}=V_i F_n(u_i)cos(t_i)$, $i=x,y$, assuming weak
modulation strengths. The odd powers of $\epsilon_{n,t_i}$ vanish
after integrating over $t_x$ and $t_y$. Then, to second order in
the modulation potential, the even powers can be easily integrated
and finally we obtain \beq \label{ap1} \frac{D(E)}{D_0}=
\frac{\hbar \omega_c}{\sqrt{\pi}}\frac{1}{\Gamma}\sum_n
e^{-(E-E_n)^2/\Gamma^2}\Big [ 1+ \frac{\Delta_n^2}{2}
\frac{2(E-E_{n})^2-\Gamma^2}{\Gamma^4}\Big ]
 \eeq where $\Delta_n^2=|V_x F_n(u_x)|^2+|V_y F_n(u_y)|^2$ and
$D_0=m^{\star}/\pi\hbar^2$ is the DOS of a free 2DEG at $B=0$.
Near the center of the $\it{n}th$ Landau level, the expression
\label{ap1}is valid only when $|V_i F_n(u_i)|^2\leq \Gamma$,
$i=x,y$.

Using the Poisson's summation formula\cite{Abra}, we carry out the
summation over $\it{n}$ and the DOS becomes
\begin{eqnarray} \label{Adosgaus}
\begin{array}{llc}
\displaystyle{\frac{D(E)}{D_0}}=1&-\frac{\displaystyle{1}}{\displaystyle{4\pi}}
\bigg(\dis{  \frac{V_x}{E} }  \bigg)^2\kappa_x
\exp{\big(-(\kappa_x\Gamma/2E)^2\big)}\Big[ \sin{(2\kappa_x)}
+ \cos{(2\kappa_x)/\kappa_x}\Big ]\\%\nonumber \\
&\\
&-\dis {\frac{1}{4\pi}\bigg(\frac{V_y}{E}\bigg)^2\kappa_y
\exp{\big(-(\kappa_y\Gamma/2E)^2\big)}\Big[ \sin{(2\kappa_y)} +
\cos{(2\kappa_y)/\kappa_y}\Big ] } \\%\nonumber \\
&\\
 &+2 \dis{ \sum  \raisebox{-2.5ex}{\hspace{-15pt} \mbox{\scriptsize
 $n$}}}   \hspace{5pt}
 (-1)^n\exp{\big(-(n\pi\Gamma/\hbar
\omega_c)^2\big)}\cos{(2\pi nE/\hbar \omega_c)}\\%\nonumber
&\\
 &\times \Big [1- \dis { \frac{2\pi n^2}{\kappa_x} \Big (\frac{V_x}
{\hbar \omega_c}\Big )^2 \cos^2{(\kappa_x-\frac{\pi}{4})}
 -\frac{2\pi n^2}{\kappa_y} \Big (\frac{V_y}{\hbar \omega_c}\Big )^2
 \cos^2{(\kappa_y-\frac{\pi}{4})}\Big]}\;
\end{array}
\end{eqnarray}
where
\begin{equation} \label{A} \kappa_x=\sqrt{\frac{2E}{\hbar
\omega_c}}\hspace{4pt}\frac{2{\pi}l}{a_x}, \hspace{10pt}
\kappa_y=\sqrt{\frac{2E}{\hbar
\omega_c}}\hspace{4pt}\frac{2{\pi}l}{a_y}
\end{equation}

Figure 1(a) shows the DOS as a function of energy. We have adopted
\begin{figure}[h!]
\label{fig2}
\begin{center}
\includegraphics[width=.6\textwidth]{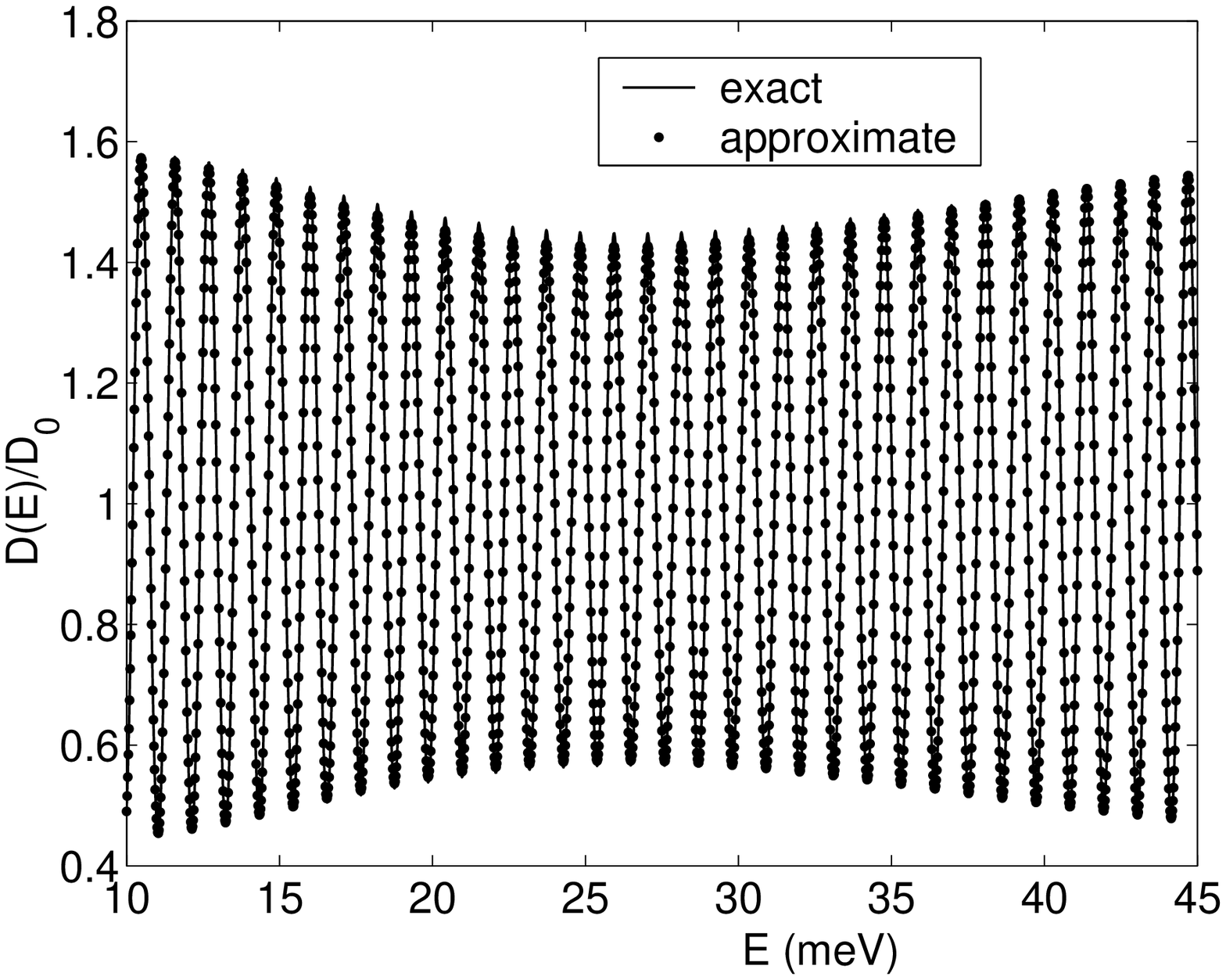}˜(a)\\
\end{center}
\begin{center}
\includegraphics[width=.6\textwidth]{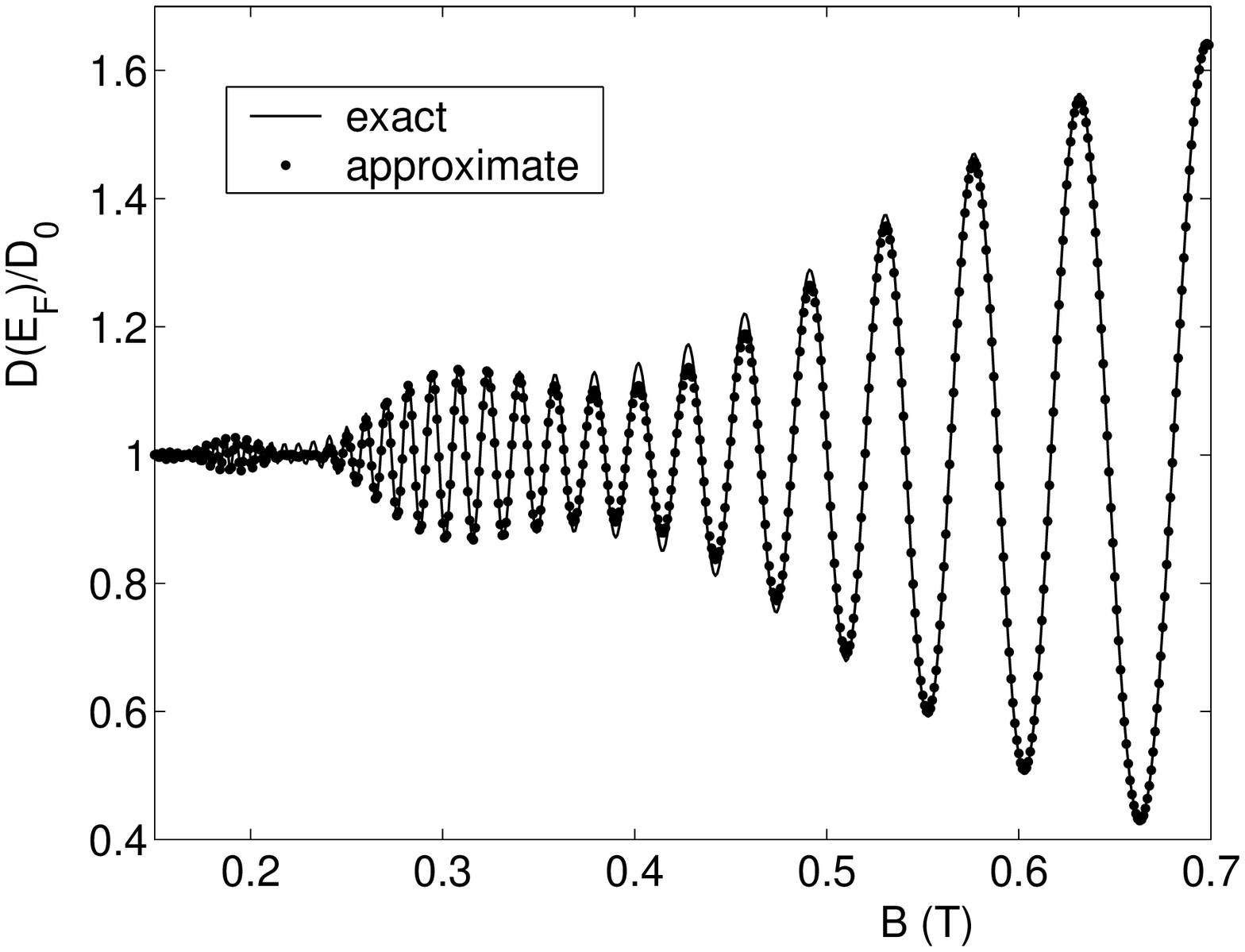}˜(b)\\
\end{center}
\caption{Calculated DOS of a 2DEG in a periodic 2D electrostatic
modulation (a) as a function of energy at $B=0.64$~T and (b) as a
function of magnetic field B. The solid curve corresponds to exact
numerical results and the dotted to approximate from
Eq.(\ref{Adosgaus}). The parameters used are $V_x=V_y=0.3$~meV,
$a_x=a_y=365$~nm, $E_F=11.43$~meV, $\Gamma=0.35\sqrt{B}$~meV.}
\label{ss3.f2}
\end{figure}
a field-dependent width $\Gamma=0.35\sqrt{B}$~meV which is
consistent with most of the earlier numerical studies in the low
magnetic field range\cite{Knob}. The other parameters used in the
numerical calculation are $V_x=V_y=0.3$~meV, $B=0.64 $~T,
$a_x=a_y=365$~nm. The solid curve shows the result from the
analytic expression (\ref{Adosgaus}) and the doted one is obtained
numerically from Eq. (\ref{dosgaus}). As it can be seen, the
agreement between the curves is good for most of the energies. In
Figure 1(b) we show the DOS at the Fermi energy as a function of
the magnetic field. The Fermi energy is evaluated for given
electron density $n_{e}$ inserting Eq. (\ref{dosdelta}) in the
relation
\begin{equation}
\label{fermi} n_e=\int_0^{\infty} dE f(E-E_F)D(E)
\end{equation}
where $f(x)=[1+e^{x/KT}]^{-1}$ is the Fermi-Dirac distribution
function. For the parameters listed above, the correction to the
Fermi energy due to weak two-dimensional modulation, is found to
be of order $10^{-3}$ over the range $B<0.7$~T and hence, for
$n_e=3.2 \times 10^{11}$~$cm^{-2}$, a fixed value of $E_F=11.43$~
meV has been adopted. Good agreement is found between the
numerical and the approximate curves except at certain magnetic
fields for which the amplitude of the DOS-oscillations is slightly
underestimated. The DOS exhibits a peak at each Landau-band
center. We observe that the weak 2D-periodic potential produces
clear modulation of the envelope of the DOS-oscillations and a
beating structure appears in the range $0.15~T <B < 0.45~T$ (Weiss
- oscillations regime). The above beating structure is not
observed in the 1D weak modulation case.

\section{Beats in the magnetocapacitance oscillations}

The capacitance of a system consisting of a metal
gate-insulator-semiconductor sandwich (e.g. gated AlGaAs/GaAs
heterostructure), depends not only on the thickness of the
insulator but also on the DOS at the semiconductor side and on the
material's parameters. If the two depletion layers interpenetrate
each other, the gate voltage $V_{G}$, is connected to the electron
density by\cite{Delag} \beq \label{gate}
V_G=\frac{eL}{\epsilon_i}n_e+\frac{E_F}{e}+K \eeq where $L$ is the
thickness of the AlGaAs-layer, $\epsilon_{i}$ is the dielectric
constant of the layer and $K$ is a constant that takes into
account fixed charges in the AlGaAs. Differentiating
Eq.(\ref{gate}) with respect to $n_e$, one obtain the total
inverse magneto-capacitance\cite{Jung} per unit area $C(B)$ at a
given temperature T:\beq \label{capacitance}
\frac{1}{C(B)}-\frac{1}{C_0}=\frac{1}{e^2D_T(0)}\Big [\
\frac{D_T(0)}{D_T(B)}-1 \Big ]\eeq where $C_0$ denotes the total
capacitance at zero magnetic field, and \beq \label{denseB}
D_T(B)=\frac{\partial n_s}{\partial E_F}=\int_0^{\infty} dE
D(E)\frac{df(E-E_F)}{dE_F}\eeq

Expression (\ref{capacitance}) is valid when a change in the gate
voltage affect only the charge in the 2DEG and the gate, leaving
intact the charge of impurities in the heterostucture. This
condition is fulfilled at low temperatures in a LSSL based on
AlGaAs/GaAs heterostructure. Following a dynamical
approach\cite{Buttiker}, an analogue expression has been recently
derived by Wang  \emph{et.al.}\cite{Wang} for the frequency
dependent electrochemical capacitance in order to study
high-frequency inductive corrections in a quantum capacitor
similar to the experiment of Gabelli \emph{et.al.}\cite{Gabelli}

\begin{figure}[htb]
\begin{center}
\begin{minipage}[t]{.45\textwidth}
\includegraphics[width=\textwidth]{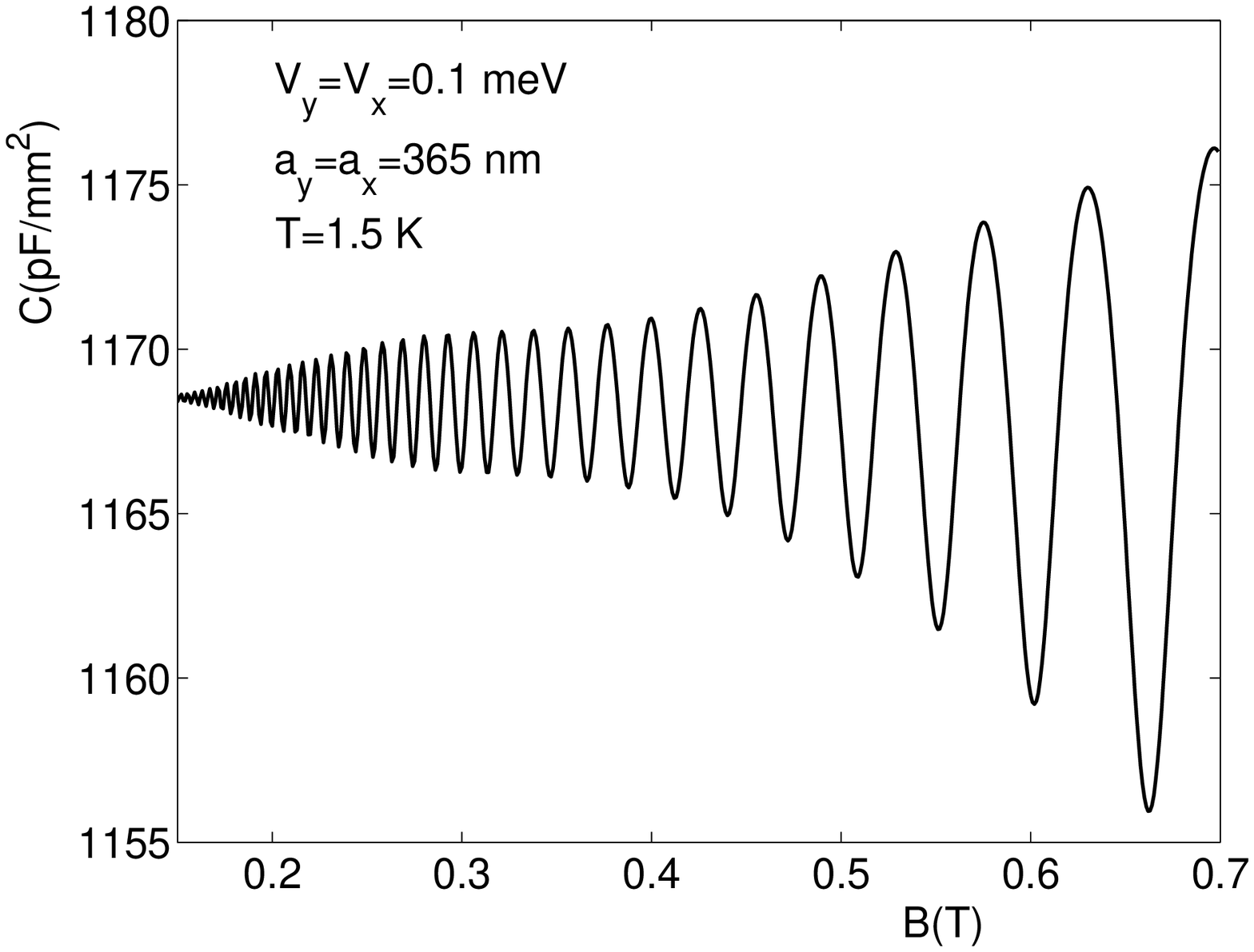}

\label{fig3}
\end{minipage}
\hfil
\begin{minipage}[t]{.45\textwidth}
\includegraphics[width=\textwidth]{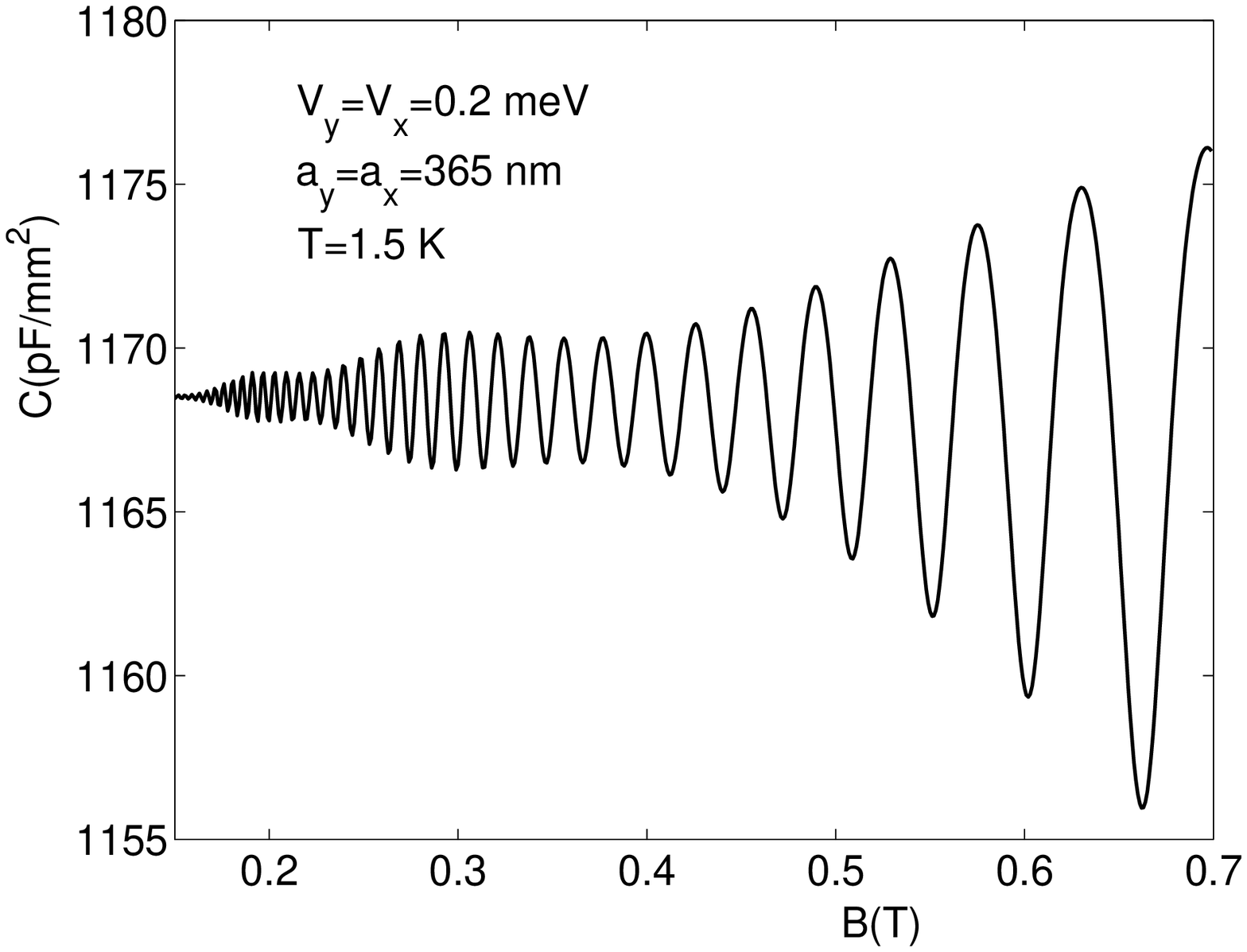}

\label{fig4}
\end{minipage}
\begin{minipage}[t]{.45\textwidth}
\includegraphics[width=\textwidth]{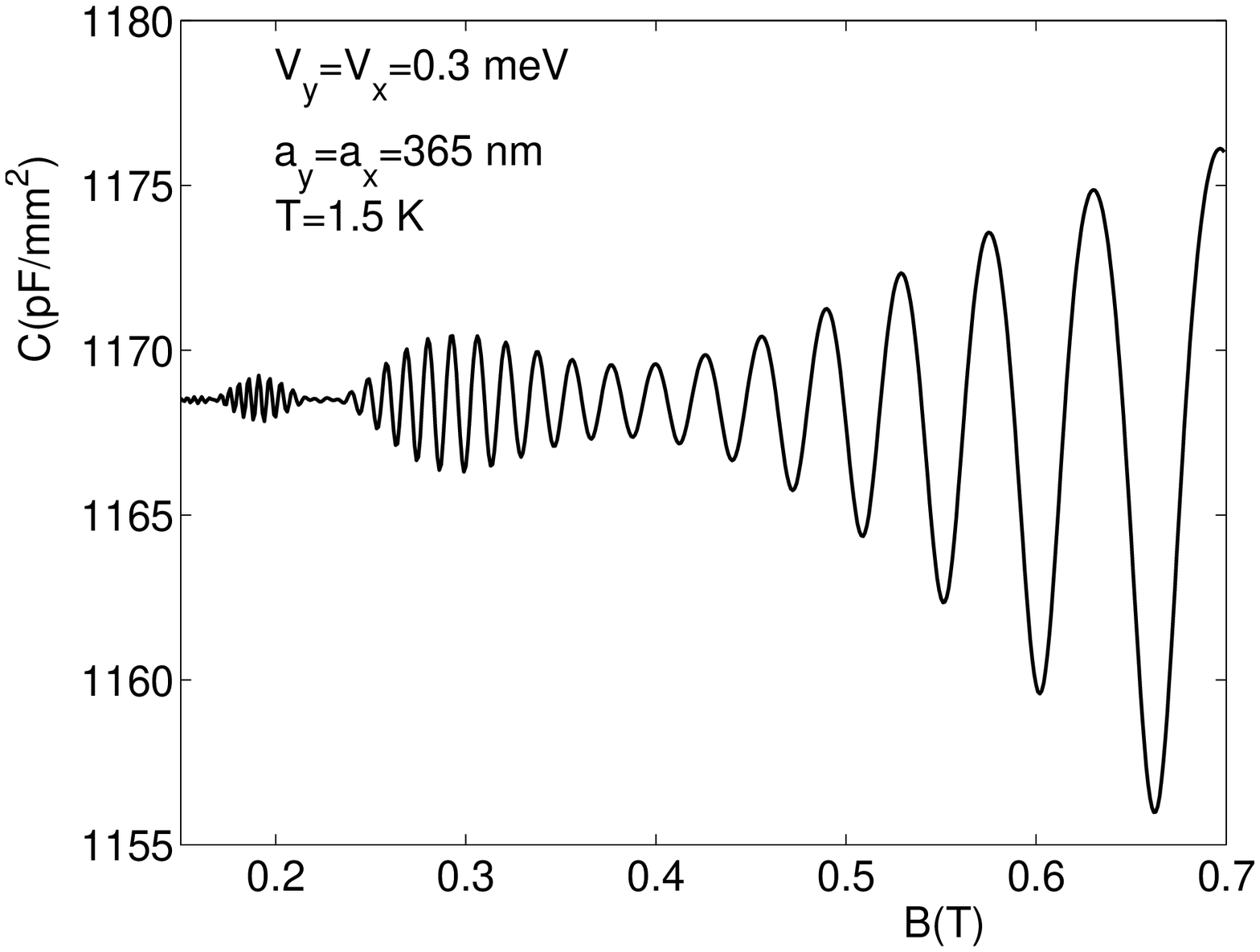}

\label{fig5}
\end{minipage}
\hfil
\begin{minipage}[t]{.45\textwidth}
\includegraphics[width=\textwidth]{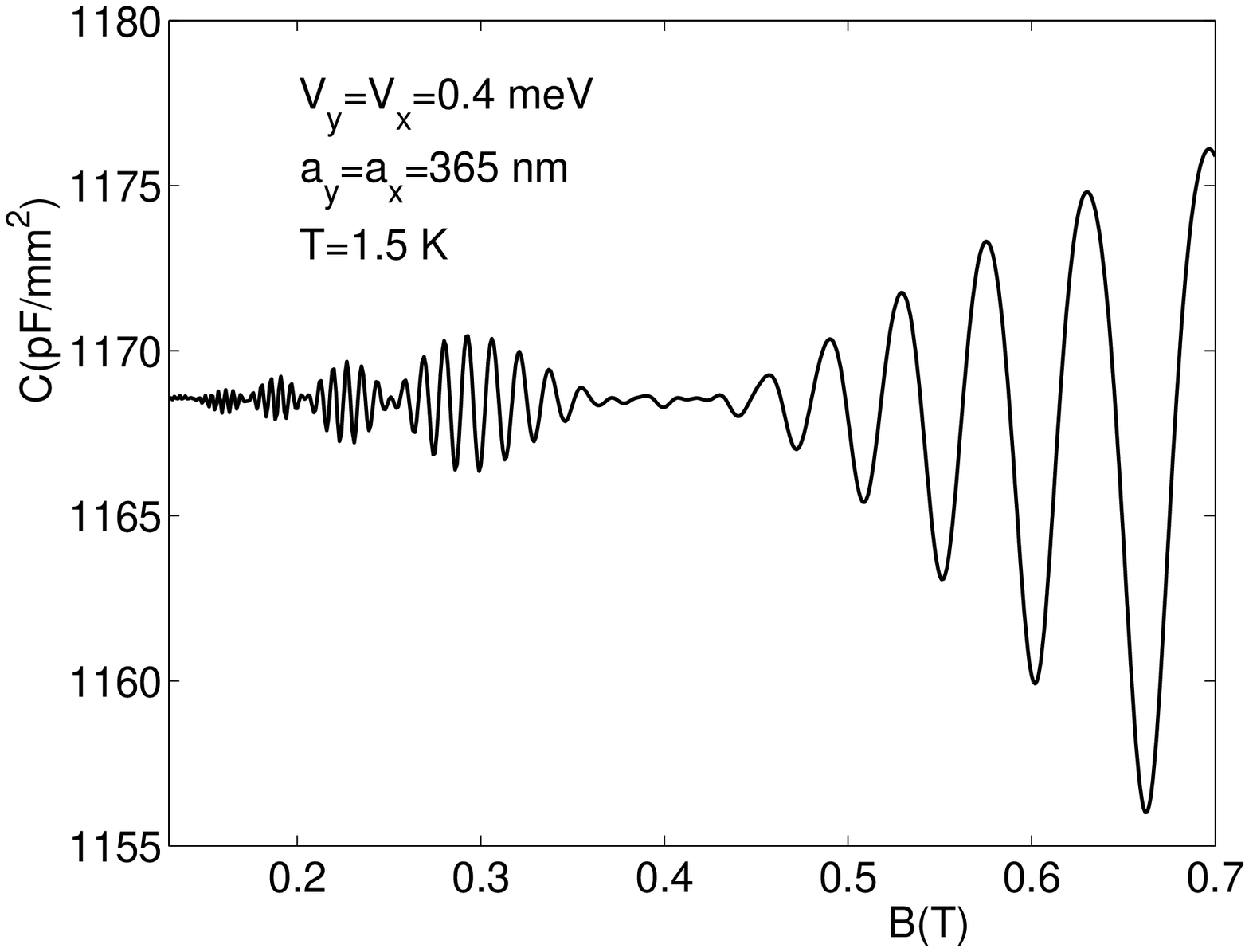}
\label{fig6}
\end{minipage}

\caption{Calculated magnetocapacitance versus magnetic field for
potential strengths $V_x=V_y=0.1,~0.2,~0.3,~0.4$~meV, scattering
broadening $\Gamma=0.35\sqrt{B}$~meV, and periods
$a_x=a_y=365$~nm.}
\end{center}
\end{figure}
Figure 2 shows our numerical results for the capacitance between
the gate and the 2DEG as a function of magnetic field at
temperature $T=1.5$~K. We consider the case of square-symmetric
modulation with $a_x=a_y=365 $~nm and
$V_x=V_y=0.1,~0.2,~0.3,~0.4$~meV. The zero-filed capacitance has
been taken from experiment\cite{Mosser} equal to $C_0=1168$
pF/~{mm}$^2$. As the modulation strength in both directions
increases, modulated magnetocapacitance oscillations with nodes (a
beating pattern) are observed in the low magnetic filed range
$0.15~T<B<0.45~T$. The phases of the oscillations changes at these
nodes and the number as well as the amplitude of beatings
increases as the modulation strength increases. We should note
that the potential strengths, are kept in the weak modulation
regime ($V_x<<E_F, V_y<<E_F$), so that Landau Level mixing is
prevented. Concerning larger magnetic fields $B>0.5$ T
(Shubnikov-de Hass oscillations range), the envelope of
magnetocapacitance maxima increases monotonically with increasing
magnetic field, while the the envelope of the magnetocapacitance
minima decreases monotonically with increasing magnetic field.

\begin{figure}[htb]
\label{fig7}
\begin{center}
\includegraphics[width=.6\textwidth]{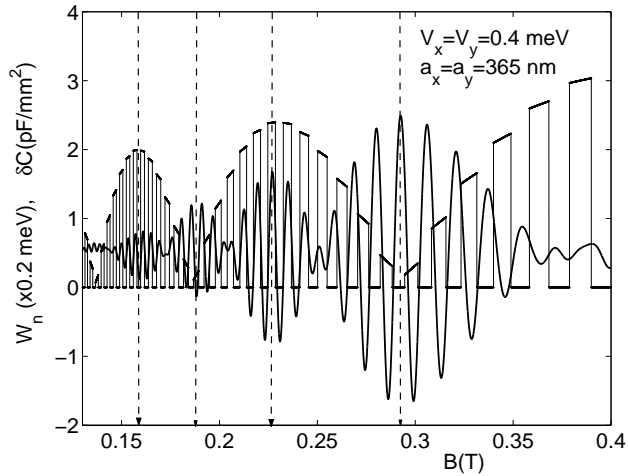}\\
\end{center}

\caption{Modulated bandwidth $W_n(B)$ and beating patterns of
magnetocapacitance oscillations for the square $LSSL$ with
$a_x=a_y=365$~nm and $V_x=V_y=0.4$~meV.}\label{fig3}
\end{figure}
The origin of the above beating structure is the oscillatory
behavior of the bi-directionally modulated bandwidth. This
behavior is clearly shown in Figure 3 where the contribution
${\delta}C(B)=C(B)-C_0$ is plotted together with the bandwidth
oscillations in the magnetic-field range of interest. The beatings
of the magnetocapacitance oscillations, at a certain magnetic
field $B$, correspond to adjacent minima or maxima of the
modulated bandwidth $W_n(B)$ and the amplitudes of the
oscillations under each envelope are closely related to the value
of $W_n(B)$. In other words, the beatings occur nearly in the
middle between adjacent `flat-band' conditions. Especially, for
the beating structure of Figure 3, we found that the cental peaks
of the beatings appear at the magnetic fields $B=0.157~
T,~0.185~T,~0.227~T,~0.292~T$. These values coincide with the
magnetic fields given by the following commensurability relation
derived from Eq.(\ref{weiss})
\begin{equation}
\label{positions}
 B_{\lambda}=\frac{2\hbar\sqrt{2 \pi n_e}}{ea(\lambda+\phi)}
\end{equation}
with $\phi=\pm 1/4$ and $\lambda=2,~3$.

\section{Concluding Remarks}

We have studied the magnetocapacitance oscillations of a 2DEG in a
LSSL under 2D weak periodic modulations at low temperatures
($T\sim1$~K). Adopting a Gaussian broadening of
magnetic-field-dependent width, an explicit analytic expression
for the oscillatory behavior of the DOS has been obtained. An
overall agreement between the exact and approximate results is
found for the relevant weak magnetic fields ($B<1$~T) and
modulation strengths ($V_x<<E_F, V_y<<E_F$). The calculated
magnetocapacitance has been shown to have a rich oscillatory
structure in this regime. As the modulation strength in both
directions increases, a beating pattern is observed in the low
magnetic field range $0.15~T<B<0.45~T$ due to the oscillatory
behavior of the bi-directionally modulated Landau-level bandwidth.
It is our aim to extend our calculations for the case of a square
LSSL with non-symmetric 2D modulation as well as the case of a
LSSL with different periods in both directions. We are not aware
of any directly relevant experimental work. We hope though that
the results described above will motivate experiments in which the
low-field magnetocapacitance could be measured in a weakly
bi-directionally modulated 2DEG of a LSSL.

\end{document}